\documentclass[a4paper,11pt]{article}
\usepackage{pos}

\title{Neutrino oscillations unlocked}
\author*[a,b]{Alexei Y Smirnov}
\affiliation[a]{Max-Planck-Institute fur Kernphysik,\\
  Saupfercheckweg 1, Heidelberg, Germany}
\affiliation[b]{International Centre for Theoretical Physics, \\
Strada Costiera 11, Trieste, Italy}

\emailAdd{smirnov@mpi-hd.mpg.de}
\abstract{
Space-time localization diagrams ``unlock" 
subtle aspects of $\nu$ oscillations such as  coherence and entanglement. 
Observability of propagation decoherence in oscillating neutrino state is discussed. 
The sizes of WPs of reactor and source neutrinos are estimated and it is shown that
damping effect due to decoherence is negligible. Among topics of interest are
transition from microscopic to macroscopic description of matter effects,
oscillations in non-linear generalization of Quantum Mechanics, influence of complex structure 
of vacuum, in particular, possible presence of cosmic strings and domain walls on  oscillations, 
neutrino oscillations driven by the effective neutrino masses generated 
via refraction on very light dark matter particles.

}

\FullConference{%
  Neutrino Oscillation Workshop-NOW2022\\
  4-11 September, 2022\\
  Rosa Marina (Ostuni, Italy)}

\begin{document}
\maketitle

\section{Introduction}
%%%%%%%%%%%%%%%%%%%%%%%%%%%%%%%%%%%%%%%%%%%%%

General idea of this talk was to give an overview of recent advances in  
the theory of neutrino oscillations as complementary to many experimental 
talks. The title was suggested by the organizers 
and I liked it so much that did not asked what ``unlocked" means. After my talk 
Eligio has clarified: that was the first NOW after COVID lock down...  

%Oscillations in vacuum
%%%%%%%%%%%%%%%%%%%%%%%%%%%%%%%%%%%

Recent studies (105 papers  with neutrino oscillation  
in titles since September 2021)  can be  classified using elements of 
standard neutrino oscillation setup: production, 
propagation, detection. 

%Fig. 1 

{\it At production} important issues are creation of coherent states, 
sizes of the neutrino wave packets,     
entanglement of neutrinos with accompanying particles.  

{\it Propagation:} Neutrino oscillations are effect of propagation in space - time. 
Neutrinos interact with VEV of scalar field(s), 
$h \langle  H \rangle $,  which produces masses and mixing. 
Interesting effects are expected if   VEV depend on coordinates: 
$\langle  H \rangle   = \langle  H(x,t) \rangle$. The coupling can be a function  
of VEV of some  field $\tau$: 
$h = h(\langle \tau \rangle)$, which  in turn depens on coordinates. 
New studies also include    
modification of geometry of space-time, metrics, oscillations in the gravitational waves 
background. 

In medium composed of particles and classical fields (e.g. magnetic fields) 
important aspects of oscillations include (i)   
transition from microscopic picture (scattering on individual electrons), 
to macroscopic one in terms of effective potentials;    
(ii) interactions of neutrinos with scalar bosons 
(DM particles) and   nature of neutrino mass; (iii) oscillations 
due to refraction:    
transition from VEV to particle densities $\langle \phi \rangle  \rightarrow \phi$;   
%%Effective mass squared  $m^2 \sim n_\phi$ 
%%z^3$ increases with decrease of $t_U$. 
(iv) treatment of oscillating neutrinos in medium  as open system, {\it etc.}

{\it At detection} features of interference and coherence were explored.  

Oscillations are the quantum mechanical effect (based on superposition and interference): 
Therefore important issues are tests of QM  and quantumness with neutrinos, 
modification of QM and therefore, evolution equation, {\it etc}.

This talk covers few aspects mentioned above:  
(i)  space-time localization diagrams; 
(ii)   coherence, entanglement and wave packets; 
(iii)  matter and  vacuum effects on propagation.

%%Other aspects-see talks by  B. Dasgupta, L. Johns, M. Blasone. 

\section{Space-time localization diagrams}
%%%%%%%%%%%%%%%%%%%%%%%%%%%%%%%%%%%%%%%%%%%%%

\begin{figure}
\centering
\includegraphics[width=0.5\linewidth]{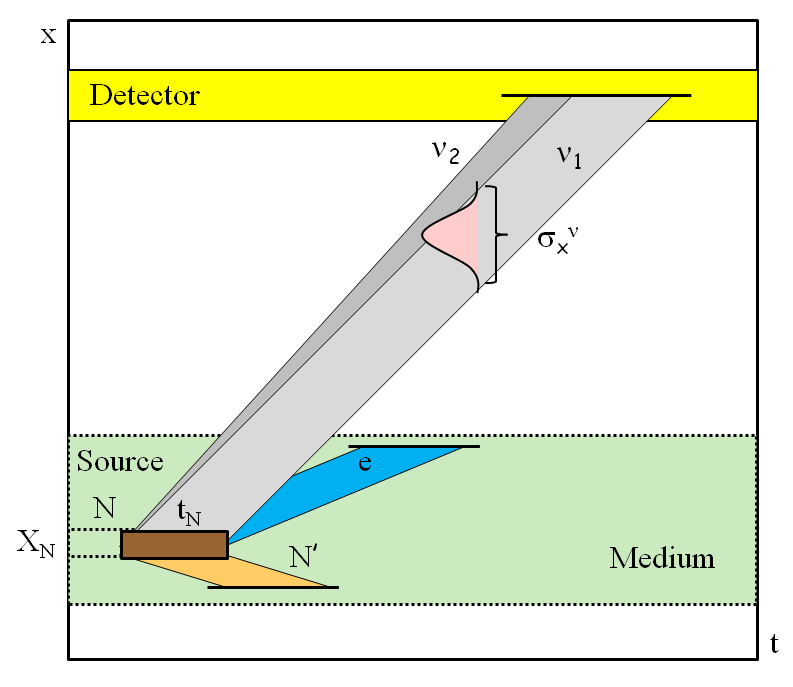}
\caption{
Space-time localization diagram for oscillations
of neutrinos produced in beta decay.
Shown are $x-t$ localization areas of the decaying 
nucleus (brown), daughter nucleus (orange), electron (blue)
and neutrino mass states (gray). The slopes are determiined 
by group velocities of particles. Yellow rectangle shows localization 
of the detection process (see details latter).}
\label{fig:diagram}
\end{figure}

{\it The diagrams}  (Fig. \ref{fig:diagram})  
reflect computations of oscillation probabilities in QFT, 
and visualize various subtle issues unlocking the underlying physics 
\cite{dani}. For simplicity we  use the $2\nu$ framework. 

The produced and then propagated neutrino state can be presented as 
\begin{equation}
|\nu^P\rangle = \psi_1^P (x - v_1 t)   |\nu_1 \rangle + \psi_2^P (x - v_2 t)|\nu_2 \rangle,  
\label{eq:propst}
\end{equation}
where 
%%\begin{equation}
%%\psi_i^P = \psi_i^P(x - v_i t)  
%%\label{eq:wpwp}
%%\end{equation}
$\psi_i^P$ are the wave packets, and $v_i$ are group velocities of mass eigenstates.   
(The production region is around $x, t = 0$).  
The detected state is 
\begin{equation}
|\nu^D\rangle = \psi_1^D (x - x_D,  t - t_D)|\nu_1 \rangle + \psi_2^D (x - x_D,  t - t_D)
|\nu_2 \rangle, 
\label{eq:detst}
\end{equation}
here 
%%\begin{equation}
%%\psi_i^D = \psi_i^P(x - x_D,  t - t_D)  
%%\label{eq:wpwp}
%%\end{equation}
$\psi_i^D$  are the detection WP with  $(x_D, t_D)$ being the  
coordinates of center of detection region. 
Amplitude of oscillations is given by  projection of the propagated state onto the 
detection state: $A (x_D, t_D) = \langle \nu^D|\nu^P \rangle$. For simplicity one can  take
\begin{equation}
\psi_i^D(x - x^D, t - t_D) = 
\delta (x - L) \psi_i^D(t - t_D),  
\label{eq:detstate}
\end{equation}
where $L$ is the baseline. Then after integration over $x$ one obtains  
\begin{equation}
 A (L, t_D) =   \sum_i  A_i (L, t_D)  = \sum_i \int dt~  
\psi_i^{D*} (t - t_D) \psi_i^P (L - v_i t). 
\label{eq:amplit}
\end{equation}
Here  $A_i (L, t_D)$ can be treated as the generalized WP
which includes localizations of both production and detection processes.  
The  oscillation probability equals
\begin{eqnarray}
P(L) & = & \int d t_D |A(L, t_D)|^2 = 
\int d t_D \left[
|A_1(L, t_D)|^2 + |A_2(L, t_D)|^2 \right] \\ 
& + &  
2Re \int d t_D A_1(L, t_D)^* A_2(L, t_D). 
\label{eq:prob}
\end{eqnarray}
%%where 
%%\begin{equation}
%%A_i(L, t_D) = \int dt \psi_i^D* (t – t_D) \psi_i^P(L – v_i t),
%%\label{eq:genwp}
%%\end{equation}
Further integration should be done
over interval  of baseline $L$ due to finite sizes of the source and detector.
In what follows we will consider
production  and detection separately and refer 
to WP as the produced and propagating WP. 
The spread of individual WP in the course of propagation is neglected.

%%%%%%%%%%%%%%%%ffff1%%%%%%%%%%%%%%%%%%%%%%%%%%%%%%%%%%%%%%%
%%%%%%%%%%%%%%%%%%%%%%%%%%%%%%%%%%%%%%%%%%%%%%%%%%%%%%%%%%%%%%%%%%%%%

{\it At the detection}  there are two extreme cases determined by time widths of the 
produced, $\sigma_t^P$, and detected,  $\sigma_t^D$, WPs.

\begin{figure}
\includegraphics[width=0.45\linewidth]{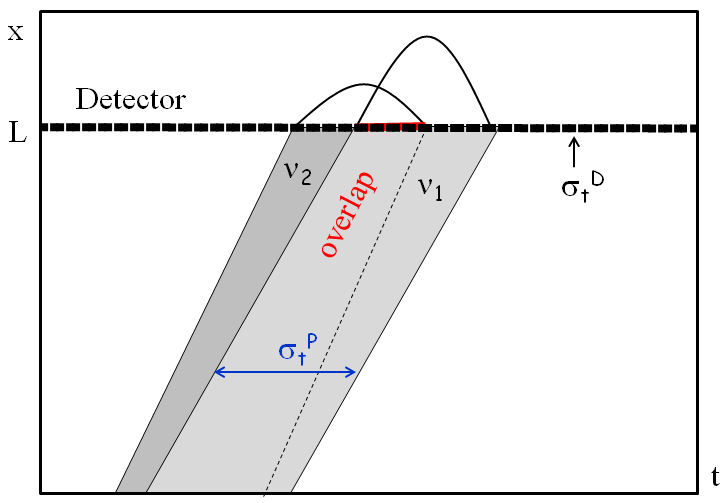}\hskip 0.5cm
\includegraphics[width=0.45\linewidth]{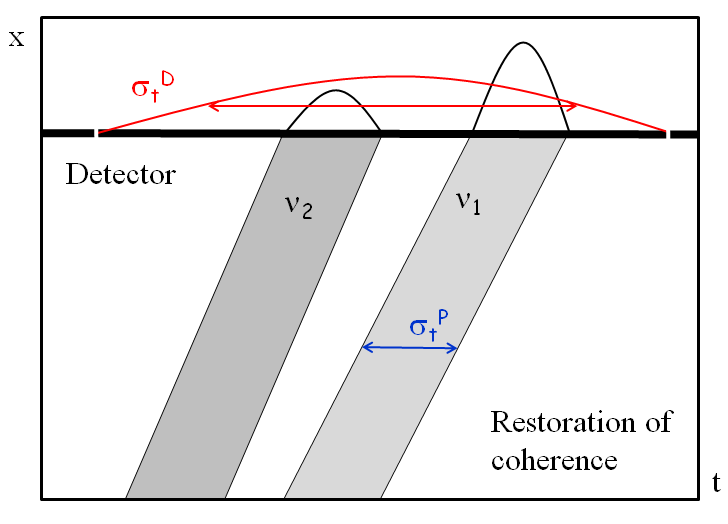}
\caption{
The $x - t$ localization of the detection process.
Black narrow rectangles show coherent areas of detection; 
grey bands show localizations of neutrino mass eigenstates. Left:
short coherence time of detection: interference and oscillations are determined by overlap of the WP
neutrino bands. Right: long coherence time of detection - 
restoration of coherence in spite of separation of WPs.}
\label{fig:wpdet}
\end{figure}

1. Short coherence time of detection $\sigma_t^D \ll \sigma_t^P$. 
In this case $\psi_i^D (t - t_D) \propto \delta(t - t_D)$, 
%$
%\psi_i^D(t - t_D) \sim \delta(t - t_D),
%$
and consequently, according to Eq.  (\ref{eq:amplit}) 
\begin{equation}
A_i(L, t_D) = \psi_i^P(L - v_i t_D).
\label{eq:short}
\end{equation}
Therefore the interference (\ref{eq:prob}) is determined by overlap of the produced WPs 
(Fig. \ref{fig:wpdet}, left).

%%%%%%%%%%%%%%%%ffff2%%%%%%%%%%%%%%%%%%%%%%%%%%%%%%%%%%%%%%%
%%%%%%%%%%%%%%%%%%%%%%%%%%%%%%%%%%%%%%%%%%%%%%%%%%%%%%%%%%%%%%%%%%%%%

2. Long coherence time of detection $\sigma_t^D \gg\sigma_t^P$. 
In this case 
\begin{equation}
A_i(L, t_D) \sim \psi_i^D(L/v_i - t_D).
\label{eq:long}
\end{equation}
The WP are separated when $t_{sep}$ - the time interval between arrived WP 
is bigger than $\sigma_t^P$. If however 
$\sigma_t^D \gg t_{sep}$, restoration 
of  coherence occurs (see Fig. \ref{fig:wpdet}, right). 
In other terms the detection process enlarges the arriving WPs so that they start to overlap. 

\begin{figure}
\includegraphics[width=0.5\linewidth]{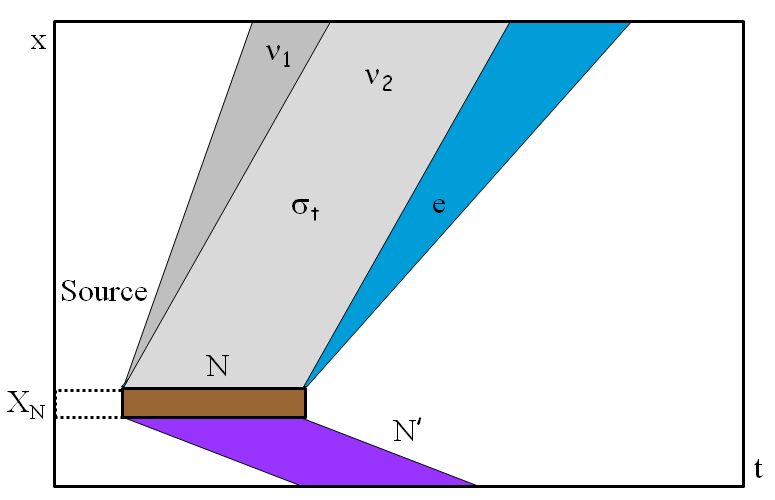} 
\includegraphics[width=0.5\linewidth]{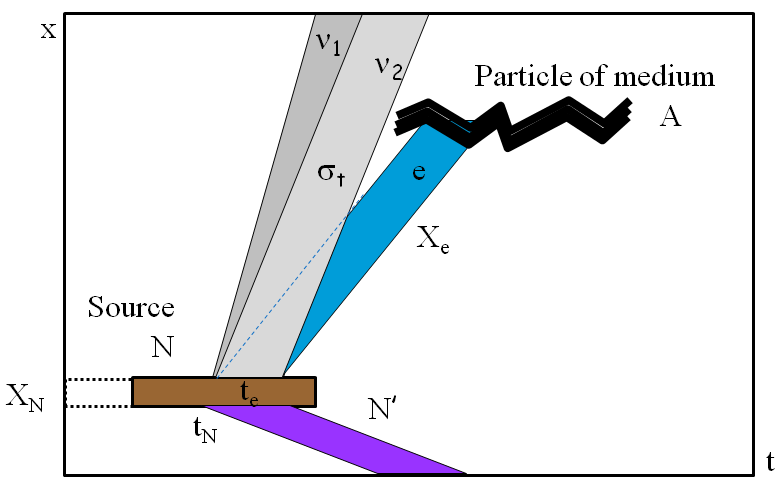} 
\caption{
The $x - t$ localization of the production process.
Left: accompanying particles (recoil and electron)
are not detected or their interactions are
negligible. Right: electron interacts with particle of medium
A and A has short mean free path (shorter than $t_N$). This localizes 
better the electron and therefore neutrino mass states. As a result,  the overlap 
becomes smaller and interference is suppressed.}
\label{fig:wpprod}
\end{figure}

{\it Production:}  
WP’s are determined by localization region of the production process: by overlap 
of localization regions of all particles involved but neutrinos. 
Consider, e.g.  the $\beta-$decay: $N \rightarrow N' + e^- + \bar{\nu}$. 
If  $N'$ and $e^-$  are not detected or 
their interactions can be neglected, 
localization of the process is given by localization of the parent nucleus $X_N$. 
The latter is determined by time between two collisions of atom contained  nucleus   
$N$, $t_N$. Then spatial size of the 
neutrino WP equals  
$$
\sigma_x \simeq v_\nu t_N \simeq X_N \frac{v_\nu}{v_N} \approx X_N \frac{c }{v_N}.
$$
Here  $c/v_N$ is huge enhancement factor, 
so that $\sigma_x \gg X_N$ (Fig. \ref{fig:wpprod}, left). 
In addition,  
one should take into account  entanglement between neutrinos and particles 
produced together with neutrinos. If $N'$ or/and $e^-$  are detected or interact, this may narrow 
the interval $t_N$,  and therefore the neutrino WP   (Fig. \ref{fig:wpprod}, right). 
If $e^-$ is detected during time interval $t_e < t_N$, the size of $\nu$ WP will be determined by $t_e$. 
If  $e^-$ interacts with particles of medium which have very short time between collisions,  $t_{coll}$, 
then $\sigma_x \simeq ct_{coll}$.
This entanglement is similar to the entanglement in the EPR paradox. 
To understand this one can consider  $\nu$ emission and interactions of $e^-$ 
as a unique process $N ~ A \rightarrow \bar{\nu} ~e^- N' A$.   
Contributions to its amplitude from different $x-t$ interactions regions of $e^-$ 
appear with random phases $\xi_k$:   
$A_{tot} = \sum_k A_e e^{i\xi_k}$ (Fig. \ref{fig:wpprod}, right),  
and therefore in the probability they will sum up incoherently.

%%%%%%%%%%%%%%%%ffff3%%%%%%%%%%%%%%%%%%%%%%%%%%%%%%%%%%%%%%%
%%%%%%%%%%%%%%%%%%%%%%%%%%%%%%%%%%%%%%%%%%%%%%%%%%%%%%%%%%%%%%%%%%%%%

%%%%%%%%%%%%%%%%%%%%%%%%%%%%%%%%%%%%%%%%%%
\section{Coherence, entanglement and wave packets }
%%%%%%%%%%%%%%%%%%%%%%%%%%%%%%%%%%%%%%%%%%%%%

In $x - t$ space separation of wave packets of mass states occurs due to difference of group velocities.
This is equivalent to integration over the energy uncertainty. 
Suppression of interference  leads to  damping of oscillations: Survival probability can be written as  
\begin{equation}
P_{ee} = \bar{P}_{ee} + 0.5 D(E,L) \sin^2\theta \cos \phi,   
\label{eq:sprob}
\end{equation}
where the damping factor $D(E,L)$ for Gaussian WP equals
$$
D(E, L) = \exp [-0.5(L/L_{coh})^2]. 
$$
Here $L_{coh}$ is the coherence length
$$
L_{coh} = \sigma_x\frac{2E^2}{\Delta m^2}. 
$$

Recall that  separation of WP leads to  
the {\it propagation decoherence},  in contrast to 
irreversible QM decoherence. In the former case the information is not lost and can be 
restored at detection. 

%13
The smaller $E$, the smaller  $L_{coh}$,  
therefore reactor neutrinos  and neutrinos from nuclear sources are most suitable  
to  search for  decoherence effects. 
The absence of damping means that 
$L \ll L_{coh}$ or explicitly 
\begin{equation}
\sigma_x \gg L  \frac{\Delta m^2}{2E^2} 
\label{eq:sigmax}
\end{equation}
giving the lower bound on $\sigma_x$.  
Analysis of  Daya Bay, RENO and KamLAND data leads to \cite{degouvea}:
\begin{equation}
\sigma_x > 2.1 \cdot 10^{-11}~ {\rm cm} ~~ (90\% ~{\rm C.L.}).      
\label{eq:sbound}
\end{equation}
Actually, this  bound corresponds to the energy resolution of the detectors $\delta_E$: $\sigma_x \sim 
1/\delta_E$.
%14
Absence of the damping due to finite momentum spread $\sigma_p$  
in Daya Bay allows to put the upper bound  \cite{an}
$
\sigma_p /p < 0.23 ~ (95\% ~{\rm C.L.})
$ 
at  $p = 3$ MeV,  which corresponds to 
$\sigma_x > 1/\sigma_E = 2.8 \cdot 10^{-11}$ cm, similar to that in (\ref{eq:sbound}).    
In future JUNO can  improve the limit down to  
$
\sigma_p /p < 0.01  (95\% {\rm C.L.})
$ 
or $\sigma_x > 2.3 \cdot 10^{-10}$ cm  \cite{wang}.  

Damping effects in the active -  sterile neutrino oscillations
were computed for various experiments 
\cite{arguelles}. 
Taking for all experiments the same value 
$\sigma_x$ (\ref{eq:sbound})  found in 
\cite{degouvea} as the  bound,  the authors arrived at the following 
conclusions:  
(i) decoherence allows to reconcile BEST result with reactor bounds; 
(ii) results of analysis of experimental data should be presented in two forms: 
with and without decoherence.  
In \cite{akh-smi} it was argued 
that these conclusion are based on incorrect value of $\sigma_x$.

%15
{\it Propagation decoherence and energy resolution:} 
Integration over the energy resolution of experimental setup 
described by $R(E_r, E)$ with width $\delta_E$ 
is  another sources of damping. 
It includes  the energy spectrum of produced neutrinos,  
or/and  energy  resolution of a detector.    
The WP of produced neutrino in the  energy representation, $f(E, \bar{E})$,  
acts on oscillations, as $R$ does, and can be attached to $R(E_r, E)$ 
\cite{akh-smi}.  In this case 
the effective resolution function can be introduced:
$$
R_{\rm eff}(E_r, E) =  \int d\bar{E} R(E_r, \bar{E})
 |f(E, \bar{E})|^2. 
$$
For Gaussian $f$ and $R$, $R_{\rm eff}$ is also Gaussian 
with width
$
\delta_E^2  + \sigma_E^2.  
$
The problem is to disentangle the two contributions.  

%16

%WP's of 

For reactor neutrinos the 
source is   $\beta$-decays 
$
N \rightarrow N' + e^- + \bar{\nu} 
$
of fragments of nuclear fission $N$ 
$N'$ quickly thermalizes and in the moment of decay
turns out to be in  equilibrium with medium with temperature $T$.  
Therefore the average velocity:
$
v_N \sim [3T/ m_N]^{-1/2}.        
$
If $N'$ and $e^-$ interactions can be neglected, 
localization of  $\nu$ production process is given by localization of $N$:         
$$
 \sigma_x \approx v_\nu t_N \approx X_N c/v_N, 
$$
where $t_N$ is the time between two essential 
(with momentum transfer $\Delta p  \geq 1 /\sigma_x$) collisions of atom containing  
$N$ with other atoms: 
$$
t_N \approx [\sigma_{AA} n_U v_N]^{-1}.                 
$$   
Here $\sigma_{AA}$ is the geometric cross-section  
$\sigma_{AA} \approx  \pi(2r_{vdW})^2$, where 
$r_{vdW}$ is the Van der Waals radius, $n_U$  
is the number density of Uranium. 
Then
\begin{equation}
\sigma_x \simeq 2.8 \cdot 10^{-3} \, {\rm cm}. 
\label{eq:sx1}
\end{equation}
%17

{\it Effect of accompanying  neutrino particles:} Its “short cut” estimation can be done taking that   
duration of $\nu$ production process equals the shortest mean free time among 
mean free times of particles involved \cite{akh-smi}. 
It was shown in \cite{akh-smi} that electrons have the shortest
time $\sigma_t = t_e = X_e/v_e$, 
where  $X_e$ is determined by ionization 
of uranium atom by $e$, $\sigma_{eU}$:
$X_e = (n_U \sigma_{eU})^{-1}$. This gives  
\begin{equation}
\sigma_x = 2 \cdot 10^{-5}~{\rm cm}, 
\label{eq:sx2}
\end{equation}
which  can be considered as the upper bound on $\sigma_x$. 

Another approach is to follow interactions of accompanying particles which determine  
their $x-t$ localizations.  
Considering a  chain of secondary  processes like $e + A \rightarrow e' + A'$ 
till thermalization of $e^-$ gives  $\sigma_x = (5 - 10)\times 10^{-5}$ cm.  

Thus,   $\sigma_x /\sigma_x^{exp} \simeq 10^5 \div 10^6$, that is, $\sigma_x \gg \sigma_x^{exp}$ 
where $\sigma_x^{exp}$ is given in (\ref{eq:sbound}).  
The corresponding energy uncertainty is 
$\sigma_E \sim 1$ eV, while energy resolution $\delta_E \sim 10^5$ eV. Therefore to be sensitive 
to WP's separation the energy resolution function should be known with better than  
$10^{-5}$ accuracy. 
For the $Cr-$source  $\sigma_x = 1.4 \cdot 10^{-5}$ cm is obtained \cite{akh-smi}.   
Thus, if additional damping is found,  it is due to some new physics and not due to WP's separation.

Large $\Delta m^2$ does not help to enhance the decoherence since oscillatory pattern shows up at $L \sim l_\nu$.
But  $L_{coh} \sim l_\nu \sim 1/\Delta m^2$ and therefore $\Delta m^2$ cancels in damping factor 
which depends on $L/L_{coh} \simeq l_\nu/L_{osc}$.   
Do situations exist in which effect of separation of WP can be tested? Among 
directions to think are  experiments 
with $L \sim L_{coh}$,   lower energies,  artificial widening of lines, e.g. by laser irradiation. 

%19.

In \cite{jones} validity of estimations of sizes of WP
(\ref{eq:sx1}, \ref{eq:sx2}), and consequently, non-unobservability 
of WP separation were questioned and reply was published in  \cite{akh-smi1}. 
In particular, it was claimed in  \cite{jones} that 
nuclear interactions inside  nucleus  measure position of parent particle (nucleon),  
and thus give much smaller sizes  of the neutrino WPs, $\sigma_x \sim 1/r_{nucleus}$, than in 
({\ref{eq:sx1}, \ref{eq:sx2}). According to \cite{akh-smi} 
WP lengths are determined  by the absolute localization of parent particle 
in the oscillation setup,  
{\it i.e.},   by the largest spatial uncertainty. The latter is given by localization 
of atom with decaying nucleus with respect to other atoms in the source. 
Notice that for reactor neutrinos de Broglie wave is much larger 
than the size of nucleus, and therefore 
neutrino is not sensitive to nuclear structure.

%21
\section{Matter, vacuum and propagation }
%%%%%%%%%%%%%%%%%%%%%%%%%%%%%%%%%%%%%%%%%%%%%

{\it From micro to macro picture.} Three different approaches were elaborated:  
(i) a coarse graining and space coordinate averaging over macroscopic 
volumes with large number of point-like scatterers  \cite{ahm1},  
(ii) summation of potentials  produced by individual scatterers on the way of neutrino
\cite{asxunjie};  (iii)  
taking into account  uncertainty in localization 
of scatterers $X_e$. For the weak interactions $r_{\rm WI}$, usually 
$X_e \gg r_{\rm WI}$, where  e.g. $X_e$  is localization of $e$ in atom \cite{fantini}.

Since $\lambda_\nu \sim  1/p_\nu \ll X_e$ 
it makes sense to consider propagation of neutrino inside individual atom with density 
profile given by wave functions of electrons. 
Modeling of medium  with individual scatterers can be done using the  castle wall profile: 
alternating layers of matter with lengths $L_a$, $L_b$ and potentials
$V_a$, $V_b$. This determines mixing angles $\theta_a$, $\theta_b$ and       
half – phases: $\phi_a$, 
$\phi_b$ acquired in the layers. 
Oscillation probability is given by \cite{akhmcastle}
$$
P = [1 - I_2/(1 - R_2 )] \sin 2 (n \zeta),  
$$ 
where $\zeta = arccos R$, $n$ -number of periods, and  
$I$, $R$ are known functions of $\phi_a, \phi_b, \theta_a, \theta_b$.  
For $\phi_a, \phi_b \ll 1$ the probability can be reduced to      
$$
P = \sin^2 2\theta_m(\bar{V}) \sin^2 0.5\phi(\bar{V}),  
$$
where
$$
\bar{V} \equiv 
\frac{V_a L_a + V_b L_b}{L_a + L_b}
$$
is nothing but the average potential, as is expected in the standard picture.  
%24

{\it WP’s and non-adiabatic evolution:}
Partially ionized (or non ionized) atoms can be considered as the electron density perturbations. 
Number density profile of electrons in atom (O, C, He) is non adiabatic \cite{kusakabe}.  
The interplay of non-adiabatic evolution and separation  (relative shift) 
of the WP’s leads to new effects:  
additional averaging of oscillations \cite{kusakabe}   
with applications to supernova neutrinos. 
No new effects is expected without WP separation 
and in the case of adiabatic evolution  \cite{kusakabe}. 
Also no new effects  appear for a profile with very sharp  (step-like) density changes 
(as in castle wall case).

However, one can  show that no  new effects 
are realized also in the case of non-adiabatic evolution of WP’s. 
Indeed, WP’s are formed at the production (at boundaries): 
$$
\psi(t, x) =  \int dp f(p) \phi_p (t,  x), 
$$
where $\phi_p (t,  x)$ are the plane waves. 
If there is no absorption or $p$-dependent interactions, $f(p)$ does not change  
in the process of evolution. Inserting $\psi$ in the evolution equation
$i d\psi/dt - H \psi = 0$ and permuting integration over $p$ and evolution we find
$$
\int dp f(p) [id \phi_p/dt - H \phi_p ] = 0. 
$$
Superposition principle and linearity of the evolution equation allow to solve first, 
equation  for $\phi_p$, and  then integrate over $p$ (form the  WPs). 
No new effects predicted in  \cite{kusakabe} are realized. In the $t-x$ space  
WP can change form in the course of evolution, 
but result integrated over time coincides with result in the $E-p$ representation \cite{porto}.
The conclusion is not clear in the case of 
$\nu \nu-$scattering when the Hamiltonian 
$H = H (\phi_p)$ leads to non-linear evolution equation.

%26
{\it Non-linear  generalization of QM}. Neutrino oscillations are the QM phenomenon, 
and therefore modifications of QM affect oscillations and inversely, oscillations can be used to 
search for deviations from canonical QM. 
In \cite{kaplan} Schrodinger equation for single particle (with Hamiltonian $H_0$) was modified as
\begin{equation}
i d\nu(t, x) /dt  =
\left[ H_0 + \epsilon  \frac{q^2}{4\pi}
\int d^4 x'
|\nu(t', x' )|^2 G_r(t, x; t',  x' ) \right] \nu(t, x).
\label{eq:nlqm}
\end{equation}
Here $G_r$ is the retarded Green function for scalar (Higgs) field $\phi$,
$q$ is charge - Yukawa coupling constant
of $\nu$ and $\phi$:  $q = m_\nu /v$.
In this framework the produced neutrino
state $\nu^{(p)}$ evolves as \cite{ghergetta}  
\begin{equation}
\nu^{(p)}(t) = U[t, t_p,\nu^{(p)}] \nu^{(p)}(t_p),  
\label{eq:evmatr}
\end{equation}
where the evolution operator $U$  depends on 
$\nu^{(p)}$ via integral in (\ref{eq:nlqm}) thus making the problem 
non-linear.  The operator  can be parametrized as
$$
U(t, t_p, \nu^{(p)}(t)) = U_0(t, t_p) + 
\epsilon U_1[t, t_p, \nu^{(p)}].   
$$
Here the first term is the standard linear evolution matrix,  while the second one is 
the first non-linear correction with $\epsilon$
being the expansion parameter. 
Consequently, the evolving  state (\ref{eq:evmatr}) 
can be written as 
$$
\nu^{(p)}(t)  = \nu^{(p,0)}(t) + 
\epsilon \nu^{(p,1)}(t).   
$$
Equation for the correction  $\nu^{(p,1)}(t)$ in coordinate representation is  
$$
id \nu^{(p,1)}(t)/dt = H_0 \nu^{(p,1)}(t) + 
G[t, x, \nu^{(p,0)}] 
$$
with the last inhomogeneous term. 
In  \cite{ghergetta} to construct $G$ the Yukawa interactions of neutrinos with scalar field 
$(\nu \nu \phi)$ used which originate from the  D5 Weinberg's operator. 
Eventually the oscillation transition probability was computed: 
$$
P = \sin^2 2\theta \left[\sin^2 \frac{1}{2}\phi  - 
\frac{\epsilon'}{4}\frac{m_1 - m_2}{m_1 + m_2}    \sin \phi \right], 
$$
where the correction 
$\epsilon' \equiv 81.5 \epsilon (m_1 + m_2)^2/v^2$ is very small. 

%28

{\it Vacuum and properties of oscillations.} It was proposed \cite{dvali} that neutrino 
vacuum condensate exists due to gravity. The order parameter is fixed by the observed values of neutrino mass    
$$
\langle \Phi_{\alpha \beta} \rangle = 
\langle \nu_\alpha^T C \nu_\beta \rangle 
\simeq \Lambda_G  = {\rm mev} - 0.1~  {\rm eV}.  
$$
Here $\nu_\alpha, \nu_\beta $  are flavor neutrino states  fixed by weak (CC) interactions 
and  charged leptons 
with mass generated by usual Higgs field.  
The condensate leads to cosmological phase transition at $T \sim \Lambda_G$ at which  
neutrinos get masses $m_{\alpha \beta} 
\simeq  \langle \Phi_{\alpha \beta} \rangle$,    
and the mass matrix can written as 
$m =   
U(\theta)^T \langle \Phi \rangle U(\theta)$, where 
$ \langle \Phi \rangle \equiv   diag 
(\Phi_{11}, \Phi_{22}, \Phi_{33})$,   
and $U(\theta)$ is the mixing matrix. 
At $T < \Lambda_G$  the relic neutrinos form bound states  $\phi = \nu_\alpha^T C \nu_\beta$, 
they  decay and annihilate into $\phi$ forming ``neutrinoless" Universe. 
The flavor symmetry of system, $SU(3) \times U(1)$, is spontaneously 
broken by the neutrino condensate with  $\phi$ 
being the Goldstone bosons. $\phi$  get small masses due explicit symmetry breaking 
by weak interactions via loops.    

%29
%%Mixing and topological defects
Symmetry breaking:  $SU(3) \rightarrow Z_2 \times  Z_2 \rightarrow I$ 
produces global topological strings in the first step and 
domain walls in the second one, thus forming 
the string-wall network \cite{dvali2}.
The length scale of strings and  
inter-string separation  equal 
\begin{equation}
 \xi \simeq 10^{14}\,  {\rm m } ( \lambda/a_G) 
(\Lambda_G/ 1{\rm meV })^{7/2},          
\label{eq:xidef}
\end{equation}
where $\lambda$ is self-coupling of string 
field $\Phi$, and $a_G$ is the scale factor of phase transition. 

Travelling  around string winds VEV 
$\langle\Phi \rangle$ by the $SU(3)$ transformation: 
\begin{equation}
\langle \Phi(\theta_s) \rangle = 
\omega(\theta_s)^T \langle \Phi \rangle
\omega(\theta_s), 
\label{eq:smixing}
\end{equation}
where $\omega(\theta_s)$ is the path 
transformation with angles  
$\theta_s = (\theta_s^{12}, \theta_s^{13},\theta_s^{23})$.          
After the path $\omega$ the lepton mixing changes as
$U = U(\theta) \omega(\theta_s)$ and   
over length $\xi$, $\theta_s  = O(1)$. 
The effect is observable: e.g. 
the solar system  moves through the frozen string-wall   background  
with $v = 230$ km/sec. For 6 years of Daya Bay operation  
the distance traveled is $d = vt = 
4 \cdot 10^{13}$ m, which is   
comparable to the  expected $\xi$ (\ref{eq:xidef}).

%30
{\it VEV or refraction on scalar DM?} Elastic forward scattering of $\nu$ on 
background scalars $\phi$ (DM) with fermionic mediator $\chi$ 
produces effective potential \cite{gechun}. 
%%shown in Fig ...
It has resonance  at  $s = m_\chi^2$. 
For $\phi$ at rest the resonance $\nu$ energy
equals
$$
E_R = \frac{m_\chi^2}{2m_\phi}. 
$$
For small $m_\phi$ the resonance is at low  
observable energies \cite{valera}. 
At 
$E \ll E_R$ the potential converges to the Wolfenstein limit. At high energies 
$E \gg E_R$ it has $1/E$ tail. 

%%%%%%%%%%%%%%%%ffff4%%%%%%%%%%%%%%%%%%%%%%%%%%%%%%%%%%%%%%%
\begin{figure}
\centering
\includegraphics[width=0.5\linewidth]{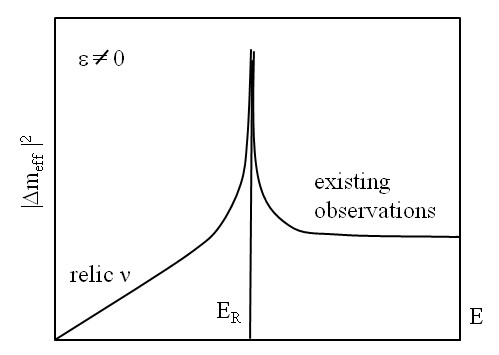}
\caption{
Dependence of the effective (refractive) neutrino mass squared on neutrino energy.}
\label{fig:effmass}
\end{figure}
%%%%%%%%%%%%%%%%%%%%%%%%%%%%%%%%%%%%%%%%%%%%%%%%%%%%%%%%%%%%%%%%%%%%%

In the Hamiltonian of propagation  the contribution of 
potential can be written similarly to 
the standard vacuum term as $\Delta m_{\rm eff}^2/ 2E $.  
Here 
$$
\Delta m_{\rm eff}^2 = 2E V 
$$
has the following limits (Fig. \ref{fig:effmass})
$$
\Delta m_{\rm eff}^2 \approx 
\frac{y^2 n_\phi}{4 m_\phi} 
\left\{\begin{matrix} 
1, \, \, \, ~~~~~~ E \gg E_R  \\
\epsilon \frac{E}{E_R}, ~~~  E \ll E_R
\end{matrix} \right. , 
$$
where $\epsilon$ is the C-asymmetry of background, 
$y$ is coupling constant of neutrino and  $\phi$, $n_\phi$ is  the number density of 
$\phi$. At high energies $\Delta m_{\rm eff}^2 = const$ 
and it has the same features as the vacuum $\Delta m^2$. Therefore in principle 
oscillations can be explained by  $\Delta m_{\rm eff}^2$, if $E_R$ is much below 
the energies at which neutrino oscillations were observed, 
that is,  much below 0.1 MeV (solar neutrinos). 

On the other hand below the resonance (Fig. \ref{fig:effmass})
$\Delta m_{\rm eff}^2$ decreases with energy. 
For $E_R = 0.01$ MeV (and $\epsilon = 1$) 
we find that in  KATRIN experiment with $E \simeq 1$ eV the effective mass 
$ m_{\rm eff} \equiv 
\sqrt{\Delta m_{\rm eff}^2} < 5 \cdot 10^{-4}$  eV,   
{\it i.e.} undetectable. 
The decrease of $m_{\rm eff}^2$  with energy allows one to avoid 
the cosmological bound on sum of neutrino masses. Indeed, 
$m_{\rm eff}^2 \propto \bar{n}_\phi \propto (1 + z)^3$, 
where $\bar{n}_\phi$ is the average density of $\phi$ 
in the Universe.  
That is, $\Delta m_{\rm eff}^2$ increased in the past, while VEV is constant. 
For relic $\nu$ at $z = 0$:  $E \simeq 10^{-4}$ eV,  
and therefore for the average  density of DM  in the Universe (the local overdensity $10^5$)  we have 
$m_{\rm eff}^2 (0) < 2.5 \cdot 10^{-16}$ eV$^2$. 
At the epoch of matter-radiation equality 
(when structures in the Universe 
started to grow) we obtain 
$m_{\rm eff}^2 (1000) \simeq  2.5 \cdot 10^{-7}$ eV$^2$ and $m_{\rm eff} (1000) \simeq  5 \cdot 10^{-4}$ eV,  
which satisfies the cosmological bound. 
Even much smaller masses are obtained in the case of $\epsilon = 0$,   
when $m_{\rm eff}^2$ decreases as $y^2$ \cite{sen}.
Existing astrophysical and laboratory bounds are satisfied, e.g., for 
$m_\chi \leq 10^{-3}$ eV, $m_\phi \leq 10^{-10}$ eV,   
$y \leq 10^{-10}$ \cite{chun}. 

\section{Conclusion}
%%%%%%%%%%%%%%%%%%%%%%%%%%%%%%%%%%%%%%%%%%%%%

Space-time localization diagrams allow to  "unlock" the key aspects of neutrino oscillations. 
Neutrino  oscillations are the tool for explorations of  properties of space and time, 
subtle aspects of QM,  fundamental symmetries  (beyond measurements of neutrino parameters). 
Effect of propagation decoherence (damping) is unobservable in the present reactor 
and source experiments. If some additional damping is found, this will be due to new physics. 

Evolution of $\nu$ state and formation of WP in the momentum space commute,  so that   
propagation decoherence is boundary (in linear case)  phenomenon.   
% (as well as production and detection decoherence). 

Effects of complex structure of vacuum, neutrino condensates, 
non-linear generalizations of QM  affect 
neutrino oscillations. Studies of neutrino oscillations allow to search for these effects.
In this connection  searches for time, space and energy dependencies of oscillation parameters 
is crucial.\\

\noindent
\section*{Acknowledgements} 

The author is thankful to Evgeny Akhmedov for numerous discussions.

%%%%%%%%%%%%%%%%%%%%%%%%%%%%%%%%%%%%%%%%%%%%%%%%%%%%%%%%%%%%%%%%%%%%%%%%%%%%%%%%%


\begin{thebibliography}{99}

\bibitem{dani} 
E.~Akhmedov, D.~Hernandez and A.~Smirnov,
%``Neutrino production coherence and oscillation experiments,''
JHEP \textbf{04} (2012), 052
%%doi:10.1007/JHEP04(2012)052
[arXiv:1201.4128 [hep-ph]].


\bibitem{degouvea}
A.~de Gouv\^ea, V.~De Romeri and C.~A.~Ternes,
%``Combined analysis of neutrino decoherence at reactor experiments,''
JHEP \textbf{06} (2021), 042
%%doi:10.1007/JHEP06(2021)042
[arXiv:2104.05806 [hep-ph]].

\bibitem{an}
F.~P.~An \textit{et al.} [Daya Bay],
%``Study of the wave packet treatment of neutrino oscillation at Daya Bay,''
Eur. Phys. J. C \textbf{77} (2017) no.9, 606
%%doi:10.1140/epjc/s10052-017-4970-y
[arXiv:1608.01661 [hep-ex]].

\bibitem{wang}
J.~Wang \textit{et al.} [JUNO],
%``Damping signatures at JUNO, a medium-baseline reactor neutrino oscillation experiment,''
JHEP \textbf{06} (2022), 062
%%doi:10.1007/JHEP06(2022)062
[arXiv:2112.14450 [hep-ex]].


\bibitem{arguelles}
C.~A.~Arg\"uelles, T.~Bert\'olez-Mart\'\i{}nez and J.~Salvado,
%``Impact of Wave Packet Separation in Low-Energy Sterile Neutrino Searches,''
[arXiv:2201.05108 [hep-ph]].



\bibitem{akh-smi}
E.~Akhmedov and A.~Y.~Smirnov,
%``Damping of neutrino oscillations, decoherence and the lengths of neutrino wave packets,''
JHEP \textbf{11} (2022), 082 
%%doi:10.1007/JHEP11(2022)082
[arXiv:2208.03736 [hep-ph]].

\bibitem{jones} 
B.~J.~P.~Jones,
%``Comment on ''Damping of neutrino oscillations, decoherence and the lengths of neutrino wave packets'',''
[arXiv:2209.00561 [hep-ph]].

\bibitem{akh-smi1}
E.~Akhmedov and A.~Y.~Smirnov,
%``Reply to ''Comment on ''Damping of neutrino oscillations, 
%decoherence and the lengths of neutrino wave packets'''',''
[arXiv:2210.01547 [hep-ph]].


\bibitem{ahm1}E.~Akhmedov,  JHEP \textbf{02} (2021), 107
%%doi:10.1007/JHEP02(2021)107
[arXiv:2010.07847 [hep-ph]].


\bibitem{asxunjie}
A.~Y.~Smirnov and X.~J.~Xu,
%``Wolfenstein potentials for neutrinos induced by ultra-light mediators,''
JHEP \textbf{12} (2019), 046
%%doi:10.1007/JHEP12(2019)046
[arXiv:1909.07505 [hep-ph]].

\bibitem{fantini}
G.~Fantini, A.~Gallo Rosso, F.~Vissani and V.~Zema,
%``Introduction to the Formalism of Neutrino Oscillations,''
Adv. Ser. Direct. High Energy Phys. \textbf{28} (2018), 37-119
%%doi:10.1142/9789813226098\_0002
[arXiv:1802.05781 [hep-ph]].

\bibitem{akhmcastle}
E.~K.~Akhmedov,
%``Parametric resonance of neutrino oscillations and passage of solar and atmospheric neutrinos through the earth,''
Nucl. Phys. B \textbf{538} (1999), 25-51
%%doi:10.1016/S0550-3213(98)00723-8
[arXiv:hep-ph/9805272 [hep-ph]].


\bibitem{kusakabe}
M.~Kusakabe,
%``Conversions of propagation eigenstates of supernova neutrinos by atomic electrons,''
[arXiv:2109.11942 [hep-ph]].

\bibitem{porto}
Y.~P.~Porto-Silva and A.~Y.~Smirnov,
%``Coherence of oscillations in matter and supernova neutrinos,''
JCAP \textbf{06} (2021), 029
%%doi:10.1088/1475-7516/2021/06/029
[arXiv:2103.10149 [hep-ph]].

\bibitem{kaplan}
D.~E.~Kaplan and S.~Rajendran,
%``Causal framework for nonlinear quantum mechanics,''
Phys. Rev. D \textbf{105} (2022) no.5, 055002
%%doi:10.1103/PhysRevD.105.055002
[arXiv:2106.10576].

\bibitem{ghergetta}
T.~Gherghetta and A.~Shkerin,
%``Out of this world neutrino oscillations,''
[arXiv:2208.10567 [hep-th]].


\bibitem{dvali}
G.~Dvali and L.~Funcke,
%``Small neutrino masses from gravitational \ensuremath{\theta}-term,''
Phys. Rev. D \textbf{93} (2016) no.11, 113002
%%doi:10.1103/PhysRevD.93.113002
[arXiv:1602.03191 [hep-ph]].


\bibitem{dvali2} 
G.~Dvali, L.~Funcke and T.~Vachaspati,
%``Time- and space-varying neutrino masses from soft topological defects,''
[arXiv:2112.02107 [hep-ph]].


\bibitem{gechun}
S.~F.~Ge and H.~Murayama,
%``Apparent CPT Violation in Neutrino Oscillation from Dark Non-Standard Interactions,''
[arXiv:1904.02518 [hep-ph]].


K.~Y.~Choi, E.~J.~Chun and J.~Kim,
%``Neutrino Oscillations in Dark Matter,''
Phys. Dark Univ. \textbf{30} (2020), 100606
%%doi:10.1016/j.dark.2020.100606
[arXiv:1909.10478].



\bibitem{valera}
A.~Y.~Smirnov and V.~B.~Valera,
%``Resonance refraction and neutrino oscillations,''
JHEP \textbf{09} (2021), 177
%%doi:10.1007/JHEP09(2021)177
[arXiv:2106.13829 [hep-ph]].

\bibitem{sen} M. Sen and A. Y. Smirnov, in preparation. 


\bibitem{chun}
K.~Y.~Choi, E.~J.~Chun and J.~Kim,
%``Dispersion of neutrinos in a medium,''
[arXiv:2012.09474 [hep-ph]].





 


\end{thebibliography}
\end{document}